\documentclass[12pt, preprint,aps,prb,superscriptaddress]{revtex4-2}
\usepackage{hyperref}
\hypersetup{
    colorlinks=true,
    linkcolor=blue,
    urlcolor=blue,
    citecolor=blue
}
\usepackage{standalone}
\usepackage{orcidlink}
\usepackage[autostyle]{csquotes}
\usepackage[utf8x]{inputenc}
\usepackage[T1]{fontenc}
\usepackage{natbib}
\usepackage{bpchem}
\usepackage{siunitx}
\usepackage{graphicx}
\usepackage{amsmath,amssymb,amstext}
\usepackage{breqn}
\usepackage{multirow}
\usepackage{float}
\usepackage{endnotes}
\usepackage{tabularx}
\usepackage{subcaption}
\usepackage{url}
\usepackage{setspace}

\makeatletter
\preto\maketitle{%
  \begingroup\lccode`~=`,
  \lowercase{\endgroup
  \let\saved@breqn@active@comma~
  \let~}\active@comma 
}
\appto\maketitle{%
  \begingroup\lccode`~=`,
  \lowercase{\endgroup
  \let~}\saved@breqn@active@comma %
}
\makeatother

\begin{document}

\title{Growth and Kerr magnetometry of Mn$_2$Au on a gold-capped Nb(001) substrate}

\author{Jendrik G\"ordes\orcidlink{0000-0003-4321-8133}}
\affiliation{Institut f\"ur Experimentalphysik, Freie Universit\"at Berlin, Arnimallee 14, 14195 Berlin, Germany}
\author{Christian Janzen\orcidlink{0000-0003-0232-2671}}
\affiliation{Institute of Physics and Center for Interdisciplinary Nanostructure Science and Technology (CINSaT), University of Kassel, Heinrich-Plett-Stra{\ss}e 40, 34132 Kassel, Germany}
\author{Arne J. Vereijken\orcidlink{0009-0001-5009-5859}}
\affiliation{Institute of Physics and Center for Interdisciplinary Nanostructure Science and Technology (CINSaT), University of Kassel, Heinrich-Plett-Stra{\ss}e 40, 34132 Kassel, Germany}
\author{Tingwei Li}
\affiliation{Institut f\"ur Experimentalphysik, Freie Universit\"at Berlin, Arnimallee 14, 14195 Berlin, Germany}
\author{Tauqir Shinwari\orcidlink{0000-0002-3876-3201}}
\affiliation{Institut f\"ur Experimentalphysik, Freie Universit\"at Berlin, Arnimallee 14, 14195 Berlin, Germany}
\author{Arno Ehresmann\orcidlink{0000-0002-0981-2289}}
\affiliation{Institute of Physics and Center for Interdisciplinary Nanostructure Science and Technology (CINSaT), University of Kassel, Heinrich-Plett-Stra{\ss}e 40, 34132 Kassel, Germany}
\author{Wolfgang Kuch\orcidlink{0000-0002-5764-4574}}
\email{Correspondence and requests for materials should be addressed to W.K. (email: kuch@physik.fu-berlin.de)}
\affiliation{Institut f\"ur Experimentalphysik, Freie Universit\"at Berlin, Arnimallee 14, 14195 Berlin, Germany}
\date{\today}                      

\renewcommand{\thefigure}{\arabic{figure}}
\renewcommand{\theequation}{\arabic{equation}}
\renewcommand{\thetable}{ \Roman{table}}
\maketitle

\newpage
\section*{Abstract}
We report on the epitaxial growth of antiferromagnetic $\mathrm{Mn_2Au}$ on a Nb(001) substrate capped with a pseudomorphic layer of gold. We observe a layer-by-layer growth by means of medium-energy electron diffraction and confirm stoichiometry and surface structure by Auger electron spectroscopy and low-energy electron diffraction. Evaporation of 15\,ML of ferromagnetic Fe on 12--17\,ML of $\mathrm{Mn_2Au}$ results in an exchange-coupled bilayer system with an exchange-bias shift that can be set by field-cooling from 400\,K. Areas with and without exchange bias, with domain sizes in the range of tens of µm, are identified by Kerr microscopy. Postannealing the sample at or above 450\,K after $\mathrm{Mn_2Au}$ layer growth decreases the amount of areas where Fe magnetically couples to $\mathrm{Mn_2Au}$. We conclude that exchange coupling to an interfacial Fe layer depends on the interface termination of $\mathrm{Mn_2Au}$. Our findings provide insight into the growth process of $\mathrm{Mn_2Au}$ and the coupling to an Fe layer. Our results point out the importance of growth, interface quality and termination on the magnetic properties of a $\mathrm{Mn_2Au}$/Fe bilayer which may help to improve material properties for spintronic applications.
\section{Introduction}
\label{Introduction}

In the last decade antiferromagnetic (AFM) materials have seen a rise in interest for spintronic applications \cite{MacDonald.2011, Jungwirth.2016, Baltz.2018, Jungfleisch.2018}. Due to their faster dynamics in the THz range \cite{Kampfrath.2011} compared to the GHz range of ferromagnetic (FM) materials as well as their robustness against perturbations from external magnetic fields and absence of stray fields, AFM are promising materials not only for fast and energy-efficient data storage but also for applications beyond storage like THz emitters and detectors or spin current generators \cite{Qiu.2021, Rongione.2023, Vaidya.2020, Li.2020, Jungwirth.2016, Zelezny.2018}. One way to achieve electric switching of the AFM order parameter, without the use of an interfacial heavy metal layer, is by using current-induced staggered Néel spin-orbit torques (SOT). So far, only $\mathrm{Mn_2Au}$ and {CuMnAs} are known to possess the required centrosymmetric lattice with locally broken inversion symmetry to generate these torques. After its theoretical prediction, several groups have demonstrated current-induced switching in these materials (even though it is not quite clear if magnetostriction or SOT-switching was observed) \cite{Wadley.2016, Bodnar.2018, Meinert.2018, Reimers.2023}. Recently, a reorientation of the AFM order parameter in $\mathrm{Mn_2Au}$ has been observed by strong exchange coupling to a FM Permalloy (Py) layer at the interface \cite{Bommanaboyena.2021, AlHamdo.2023}. 
\\
$\mathrm{Mn_2Au}$ is a metallic collinear AFM with a bulk Néel temperature above 1000\,K \cite{Barthem.2013} and a body-centered tetragonal crystal structure with a = 3.32\,\r{A} and c = 8.53\,\r{A}  as lattice parameters \cite{Bommanaboyena.2020}. For layers grown in the (001) plane, magnetic easy axes are parallel to the [110] directions. It has been shown that the quality of the layer growth is essential for its (magnetic) properties \cite{Sapozhnik.2019, Bommanaboyena.2020} and that the Néel temperature can be lowered in proximity to Fe \cite{Wu.2012}. So far, not many substrates are available for the epitaxial growth of $\mathrm{Mn_2Au}$(001) layers, often relying on a sputter-deposited buffer layer, like Ta, to accommodate the large lattice mismatch to the substrate \cite{Jourdan.2015, Meinert.2018, Bommanaboyena.2020, Gelen.2024}. For a well-ordered and defect-free layer-by-layer growth of $\mathrm{Mn_2Au}$(001), it would thus be ideal to use a single crystalline substrate with similar lattice parameters as $\mathrm{Mn_2Au}$. 
\\
In this paper, we report on the epitaxial growth of $\mathrm{Mn_2Au}$ directly on a metallic Nb(001) substrate capped with a pseudomorphic layer of Au. Nb, like Ta, has a very close lattice match to $\mathrm{Mn_2Au}$ with a bcc lattice and a lattice constant of a = 3.30\,\r{A}. To our knowledge, this kind of single-crystal substrate has not been used before for the growth of $\mathrm{Mn_2Au}$. Medium-energy electron diffraction (MEED) reveals a layer-by-layer growth of $\mathrm{Mn_2Au}$. After evaporation of 15 monolayers (ML) of ferromagnetic Fe on top of the $\mathrm{Mn_2Au}$ films, a two-step magnetization reversal is observed in longitudinal magneto-optical Kerr effect (L-MOKE) measurements, where the second step exhibits an exchange-bias (EB) shift of the magnetization loop as well as a temperature-dependent coercivity, indicating a coupling between FM Fe and AFM $\mathrm{Mn_2Au}$. We conclude that the first step of the magnetization reversal belongs to areas of the sample where Fe and $\mathrm{Mn_2Au}$ do not couple at the interface, while the second step belongs to areas that do couple. Postannealing the $\mathrm{Mn_2Au}$ layer to $\geq$ 450\,K, prior to Fe evaporation, reduces the coupling areas while simultaneously increasing their EB and coercivity. Utilizing Kerr microscopy, we identify coupled and uncoupled areas on the sample as well as magnetization reversal mechanisms. Our findings highlight the feasibility to expand the range of available substrates to metal single crystals for the growth of $\mathrm{Mn_2Au}$ layers as well as a better understanding of the coupling between $\mathrm{Mn_2Au}$ and Fe at the interface.

\section{Experimental details}
\label{Experiment}
\begin{figure}[htbp]
\includegraphics[width=0.35\columnwidth]{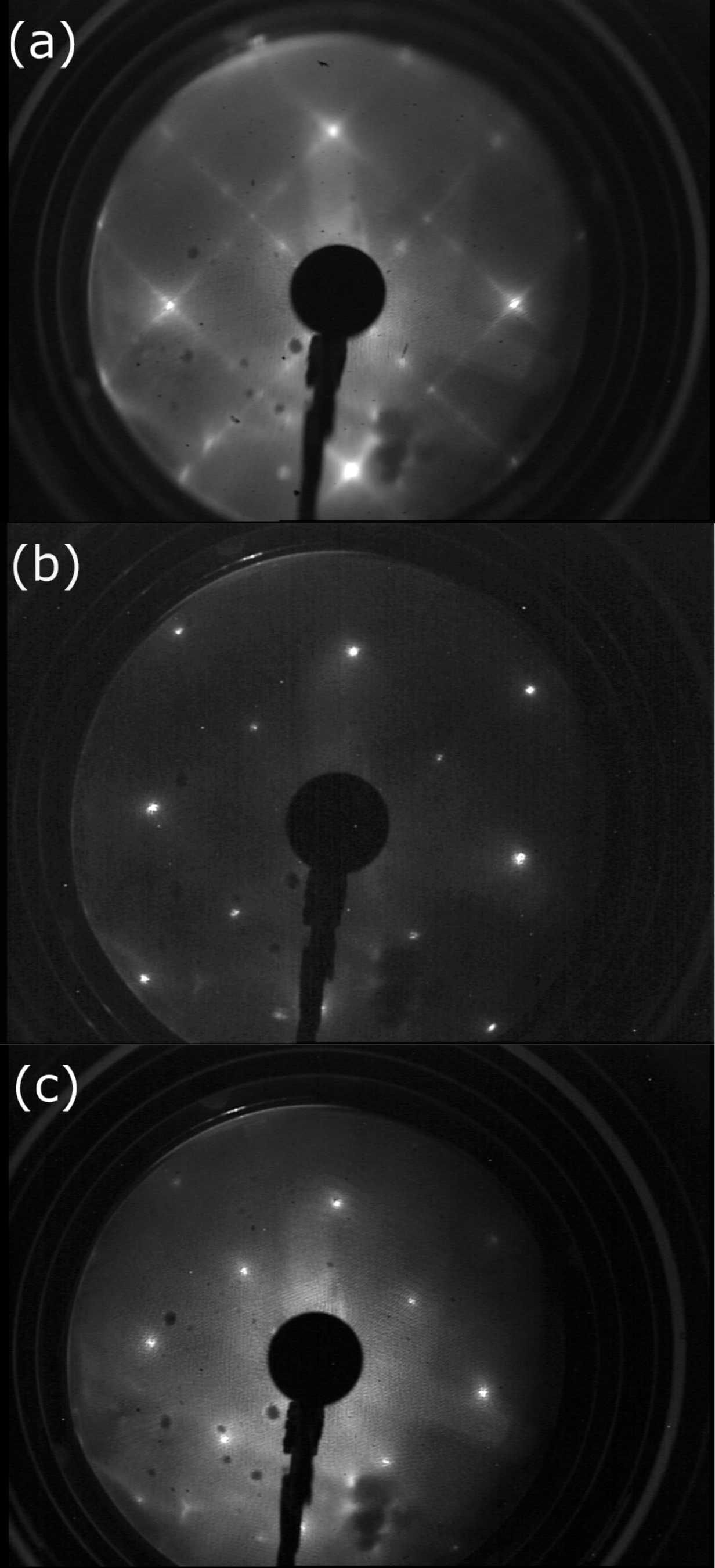}
	\centering
	\caption{LEED images of surface-oxidized Nb (a), after flash annealing of Au (b) and after evaporation of 15\,ML of $\mathrm{Mn_2Au}$ (c) at electron energies of 140--148\,eV. Additional LEED spots in (a) indicate an oxygen superstructure while (b) and (c) show clear (001)-spots corresponding to a flat surface with long-range crystallographic order.}	
	\label{leed}
\end{figure}

Samples were grown on a Nb(001) substrate with molecular beam epitaxy in UHV. The base pressure of the chamber was around $7 \times 10^{-10}$\,mbar while the pressure was kept in the lower $10^{-9}$ range during (co-)evaporation of Mn, Au, and Fe.
\\
The layer system of our samples consists of Nb(001)/NbAu/12--17\,ML $\mathrm{Mn_2Au}$/$t_F$ Fe with the Nb(001) single crystal being capped by one ML of Au which will be explained in the following. Achieving an oxygen-free Nb surface is quite challenging, requiring repeated flash annealing at temperatures close to the melting point of Nb with excellent chamber pressure conditions \cite{Farrell.1973, Usami.1977, An.2003}. Still, oxygen may segregate from the bulk to the surface, contaminating the surface and changing the lattice structure. To avoid this, we employed the substrate-cleaning protocol from Hüger \textit{et al}. \cite{Huger.2005}. The Nb(001) single crystal was first cleaned by repeated Ar+ sputtering and flash annealing at around 1900\,K for a few minutes. To prevent oxygen segregation, 10\,ML of Au were evaporated on Nb and subsequently flash-annealed to 1200-1300\,K for 70\,s. At these parameters, all the Au in excess of 1\,ML diffuses into the bulk of the Nb but still stays close enough to the surface \cite{Huger.2005}. This results in the formation of a surface-near Au-Nb alloy, which acts as a barrier for oxygen segregation, leading to an inert, flat surface with metallic character while still preserving Nb lattice parameters. Auger electron spectroscopy (AES) at 3\,keV confirmed successful flash annealing of Au by monitoring the presence of Nb and Au peaks and the absence of C and O contamination. Additionally, low-energy electron diffraction (LEED) was employed to check for a clean, flat surface with (001)-orientation without oxygen superstructure. LEED images, taken at electron energies of 140--148\,eV, are shown in Fig.\ \ref{leed} after flash annealing Nb(001) at 1900\,K (a) and after flash annealing of 10\,ML of Au (b) on Nb(001). In (a), additional LEED spots indicate an oxygen superstructure on the Nb(001) surface, which was confirmed by AES measurements, while in (b) clear (001)-spots correspond to a flat surface with long-range crystallographic order.
\\
Deposition of $\mathrm{Mn_2Au}$ was achieved at 380-390\,K (or 350\,K) substrate temperature by co-evaporation of Mn flakes in a Ta-crucible and Au from a rod wrapped in a W filament. During evaporation, a quartz crystal microbalance (QCM) was used to monitor the evaporation rates of Mn and Au, while medium-energy electron diffraction (MEED) oscillations were used to observe layer-by-layer growth of $\mathrm{Mn_2Au}$ in-situ. 
\\
After evaporation, the atomic composition of the deposited $\mathrm{Mn_2Au}$ layer was checked by AES. Surface quality, its lattice orientation and lattice parameters were checked by LEED. Fig.\ \ref{leed} (c) shows a LEED image of the sample after evaporation of 15\,ML of $\mathrm{Mn_2Au}$. The intensity of the LEED spots is reduced compared to the Au/Nb(001)-like surface, however, clear (001)-spots can be seen at the same positions. Thus, the horizontal lattice parameters of the grown $\mathrm{Mn_2Au}$ layer match the ones of the Nb(001)/NbAu substrate.
\\
15\,ML of Fe were subsequently evaporated from a rod with the sample kept at room temperature and its magnetic properties were checked by measuring the magnetization loop via in-situ longitudinal magneto-optic Kerr effect (L-MOKE) magnetometry. The azimuth of the Nb(001) single crystal was turned by 45° to apply the external magnetic field parallel to the [110]-direction of the Mn spins in the $\mathrm{Mn_2Au}$ layer. For the field-cooling (FC) process, the sample temperature was increased to 400\,K for 10\,min and an external magnetic field of 100\,mT was applied in [110]-direction before cooling down with liquid helium to 40--50\,K to set the direction of the EB. 
\\
For the ex-situ Kerr microscopy measurements, the sample was capped with at least 15\,ML of Cu before taking it out into air to prevent oxidation of Fe and $\mathrm{Mn_2Au}$. Magnetic domains were imaged by a wide-field Kerr microscope from Evico Magnetics arranged in the Köhler illumination scheme \cite{Soldatov.2017,Soldatov.2017b}. Images were drift-corrected afterwards by a software-based approach according to \cite{Akhundzada.2023} to enhance domain contrast. 

\section{Results}
\label{Results} 
\begin{figure}[htbp]
\includegraphics[width=0.55\columnwidth]{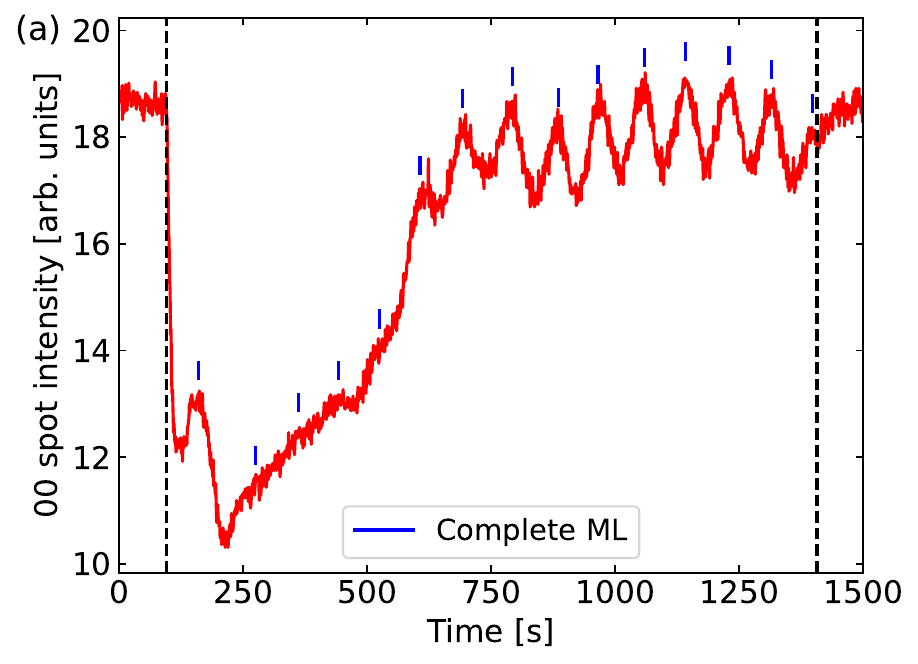}
\includegraphics[width=0.55\columnwidth]{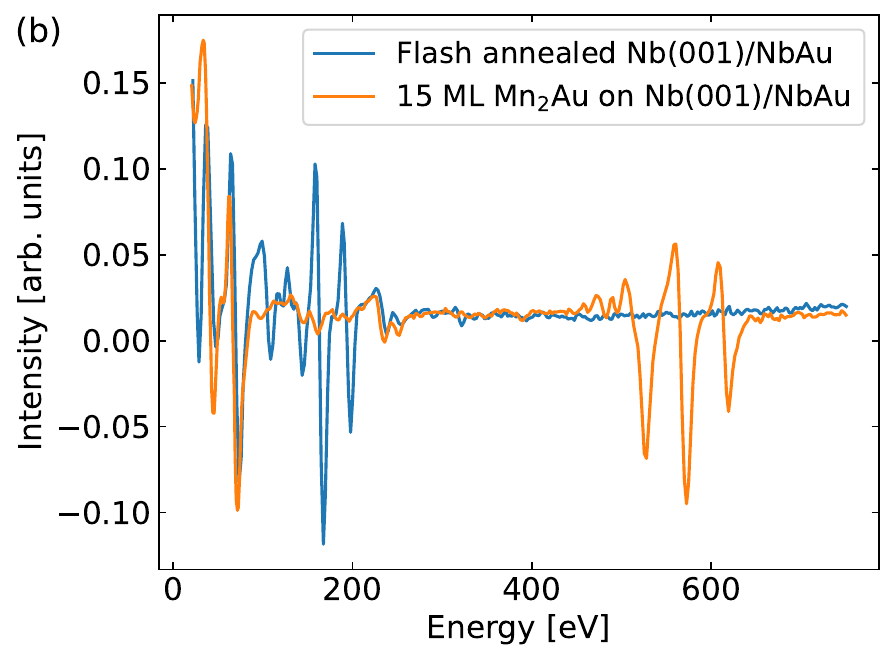}
	\centering
	\caption{(a): MEED oscillations measured in-situ during deposition. Black dotted lines indicate start and end of deposition. Blue dashes indicate a complete ML, with the second and third ML marked at likely positions interpolated from the other peaks. (b): Auger electron spectrum of the clean flash-annealed Nb(001)/NbAu substrate and after deposition of 15\,ML of $\mathrm{Mn_2Au}$.}	
	\label{MEED oscillations and AES}
\end{figure}

Fig.\ \ref{MEED oscillations and AES} (a) shows the intensity of the (00)-MEED spot during evaporation of 15\,ML of $\mathrm{Mn_2Au}$. The intensity first decreases with a local maximum (marked by a blue line) after deposition of 1\,ML. Between 1.5 and 7\,ML, the intensity of the (00)-spot increases to approximately the initial value. Strong oscillations after 6\,ML indicate a well-ordered layer-by-layer epitaxial growth. Surface roughness of the finished layer appears to be minimal. Maxima of the MEED oscillations are in good agreement with multiples of the deposited amount measured by the QCM. 
\\
The Auger electron spectrum of 15\,ML of $\mathrm{Mn_2Au}$ on Nb(001)/NbAu is displayed in Fig.\ \ref{MEED oscillations and AES} (b). Calculating the ratio of the Auger peaks (see Supplemental Material \cite{Supplement.2026}) yields a composition of $\mathrm{Mn_{67}Au_{33}}$ (67:33) indicating a deposition of $\mathrm{Mn_2Au}$ in a 2:1 ratio. Deposition of a $\mathrm{Mn_{1-x}Au_x}$ layer with a different composition than 2:1 results in a lower intensity of the MEED spots and no or only a few weak oscillations, since the lattice parameters do not fit to the Nb substrate. 
\\
LEED images, taken at electron energies of 140--148\,eV both for the clean Au/Nb(001) substrate (Fig.\ \ref{leed}(b)) and after deposition of 15\,ML of $\mathrm{Mn_2Au}$ (Fig.\ \ref{leed} (c)) show clear (001)-spots at similar positions, though the intensity of the LEED spots is slightly decreased and a bit more diffuse after $\mathrm{Mn_2Au}$ deposition. This is probably due to an increased surface roughness or reduced crystallinity throughout the whole layer.

\begin{figure}[htbp]
\includegraphics[width=0.5\columnwidth]{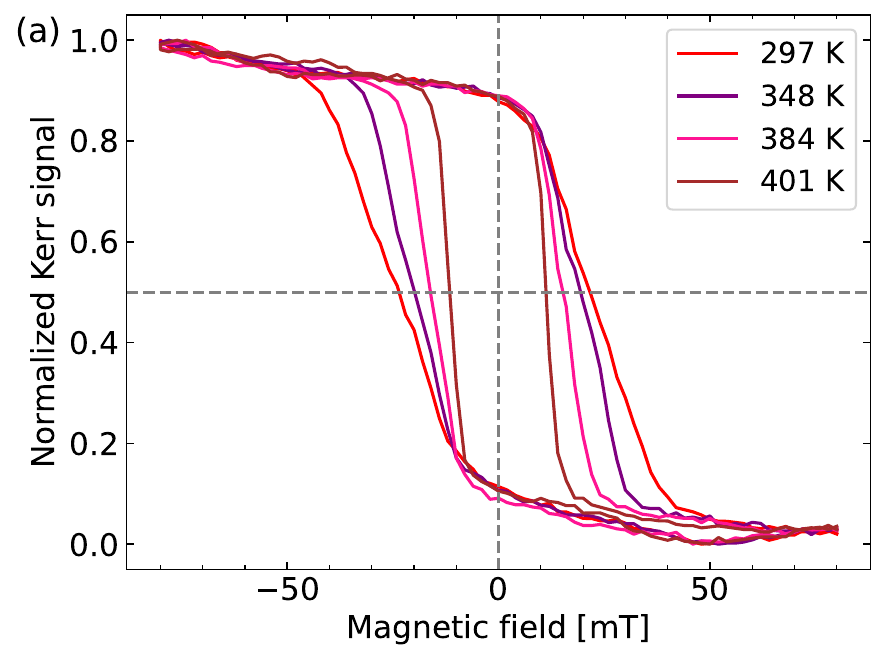}
\includegraphics[width=0.5\columnwidth]{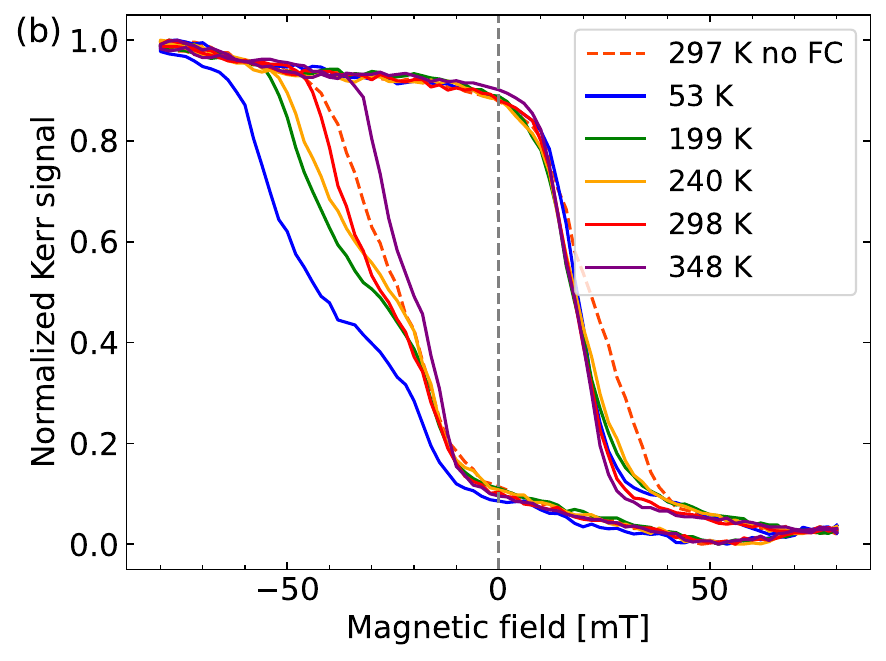}
\includegraphics[width=0.5\columnwidth]{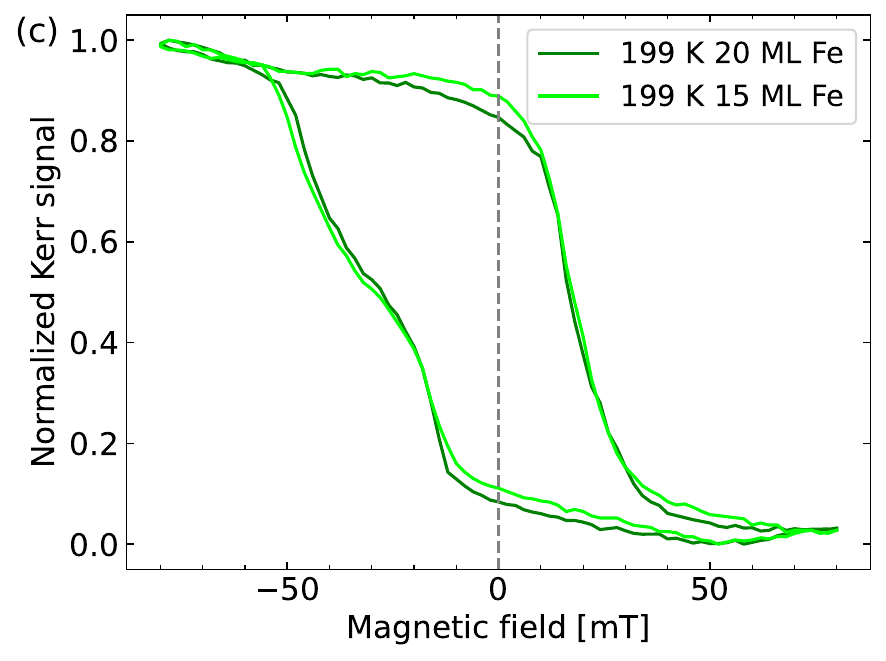}
	\centering
	\caption{Temperature dependence of hysteresis loops of Nb(001)/NbAu/15\,ML $\mathrm{Mn_2Au}$/15\,ML Fe before (a) and after (b) FC, together with the original hysteresis loop measured at RT for reference. (c) Comparison of hysteresis loops between 15\,ML and 20\,ML of Fe.}	
	\label{temperature}
\end{figure}

Fig. \ref{temperature} shows magnetization loops measured by in-situ L-MOKE after the deposition of 15\,ML of Fe. While heating up from room temperature to 400\,K, shown in Fig. \ref{temperature} (a), the coercivity of the Fe layer decreases from 22.5\,mT at room temperature down to about 11\,mT at 400\,K. This temperature-dependent coercivity suggests a coupling of FM Fe to the AFM $\mathrm{Mn_2Au}$. Since no FC process was initiated so far, the Néel vector orientation in the AFM domains is randomly distributed and averages out within the beam spot. Hence, no EB shift can be measured. 

After a FC process at 400\,K down to about 50\,K, magnetization loops were measured at different temperatures while heating back up to 348\,K and are depicted in Fig. \ref{temperature} (b). Now, one part of the hysteresis loop is shifted horizontally towards larger negative fields, in opposite direction to the setting field during the FC process. A second part of the hysteresis loop is not shifted, which results overall in a two-step magnetization reversal. Here, the coercivity of the non-shifted part of around 12.5\,mT is comparable to the one of a pure Fe layer on Nb(001). Looking at the temperature dependence, the non-shifted part of the hysteresis loop shows no change with temperature, while both the coercivity and the horizontal shift of the shifted part decrease with increasing temperature. The absolute horizontal shift decreases from around 16.4\,mT at 50\,K to 7.6\,mT at room temperature and goes to zero between 348\,K and 400\,K. The coercivity decreases from 39.0\,mT at 50\,K to 24.9\,mT at 348\,K. From these measurements we conjecture that the evaporated $\mathrm{Mn_2Au}$ layer is antiferromagnetic and we were able to set, at least partially, the direction of its moments at the interface to Fe by field cooling from 400\,K, resulting in an EB shift and increased coercivity of the shifted part. 

For 15\,ML of $\mathrm{Mn_2Au}$ the ordering temperature appears to be between 348\,K and 400\,K, since otherwise the EB could not be set.  This value is similar to the results from Wu \textit{et al}. \cite{Wu.2012} for 10\,nm of $\mathrm{Mn_2Au}$(101) grown on Fe/MgO(001). That means the Néel temperature of $\mathrm{Mn_2Au}$ is significantly reduced in the thin film compared to the bulk, where it should be above 1000\,K. Reasons for this could be the finite-size effect as well as the interface and exchange interaction to the Fe layer \cite{Wu.2012, Ambrose.1996, Won.2005, Choo.2007, Qiu.2008}.

Aside from the ordering temperature, the EB shift and increased coercivity of the shifted part, as well as their temperature dependence, are further indications for a magnetic coupling between the FM Fe and the AFM $\mathrm{Mn_2Au}$. Consequently, there are two possibilities for the two step remagnetization: (1) only the interfacial Fe moments magnetically couple to the $\mathrm{Mn_2Au}$ moments while the bulk Fe moments do not couple, (2) there are coupling and non-coupling areas between Fe and $\mathrm{Mn_2Au}$ laterally on the sample. To test these two hypotheses, we evaporated an additional 5\,ML of Fe on top for a total of 20\,ML of Fe. The comparison between the L-MOKE hysteresis loops of Nb(001)/NbAu/15\,ML $\mathrm{Mn_2Au}$/15\,ML Fe and 20\,ML of Fe is presented in Fig.\ \ref{temperature} (c). As a result, the non-coupling part remained unchanged, while the horizontal shift (EB) of the coupling part decreased. This reduction in EB is likely due to its inverse dependence on FM layer thickness. The relative amplitude, defined as the ratio of the amplitudes of the shifted part and the non-shifted part in the L-MOKE signal, did not change, which means that we can exclude that they belong to interface and bulk Fe. Instead, this confirms the presence of coupling and non-coupling areas on the sample.

\begin{figure}[htbp]
\includegraphics[width=0.49\columnwidth]{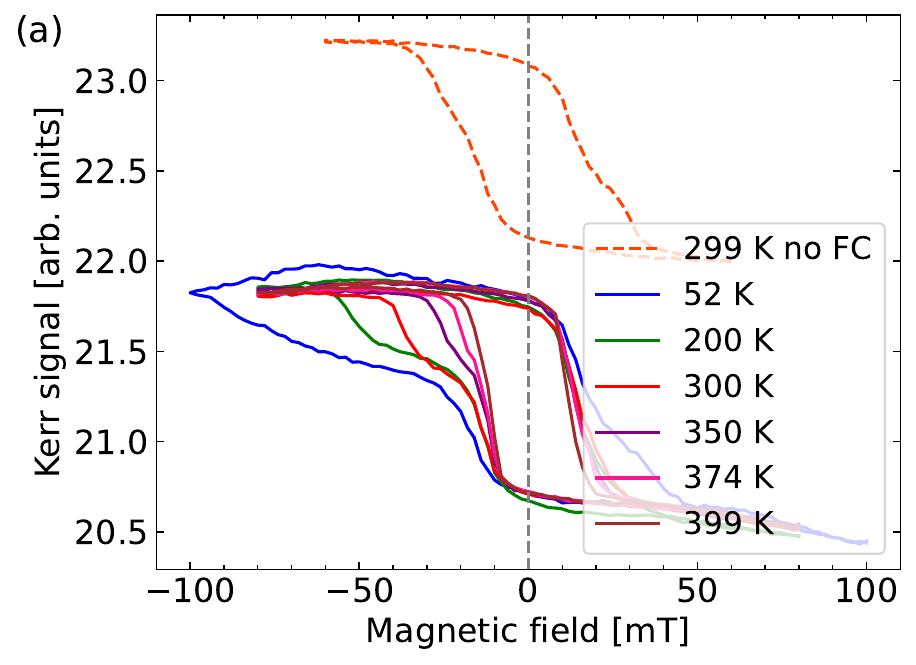}
\includegraphics[width=0.49\columnwidth]{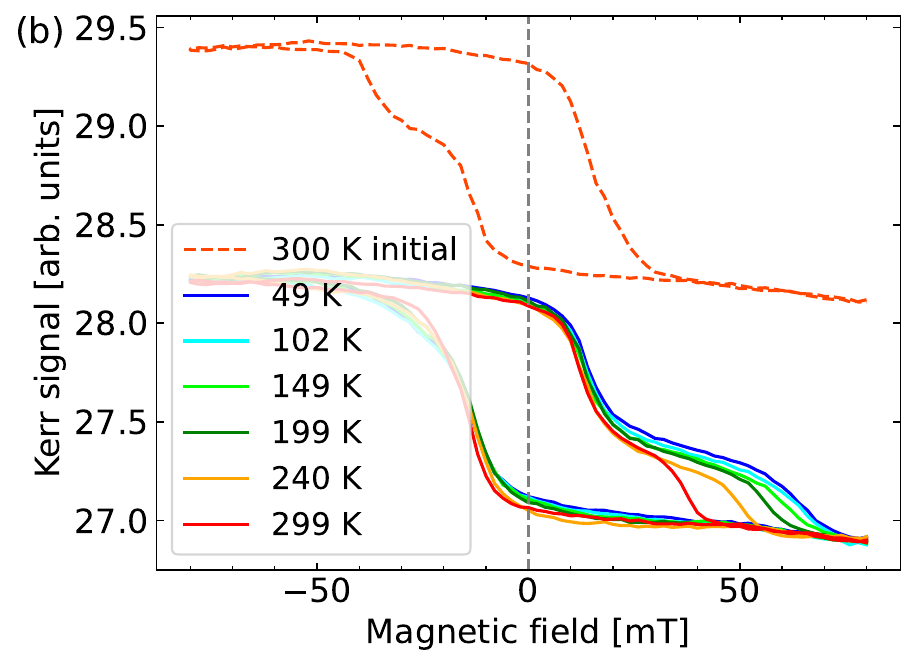}
\includegraphics[width=0.49\columnwidth]{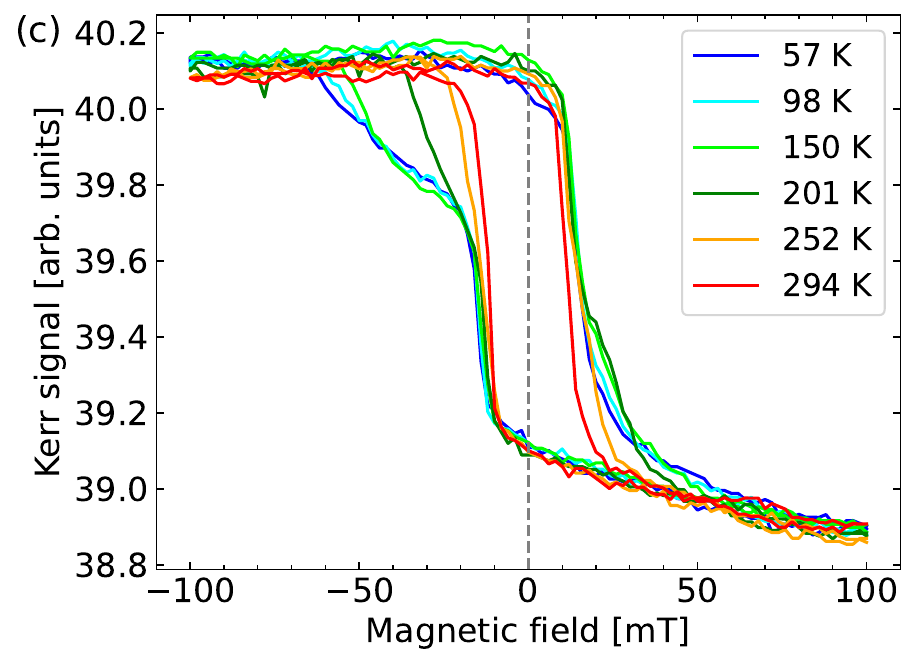}
\includegraphics[width=0.49\columnwidth]{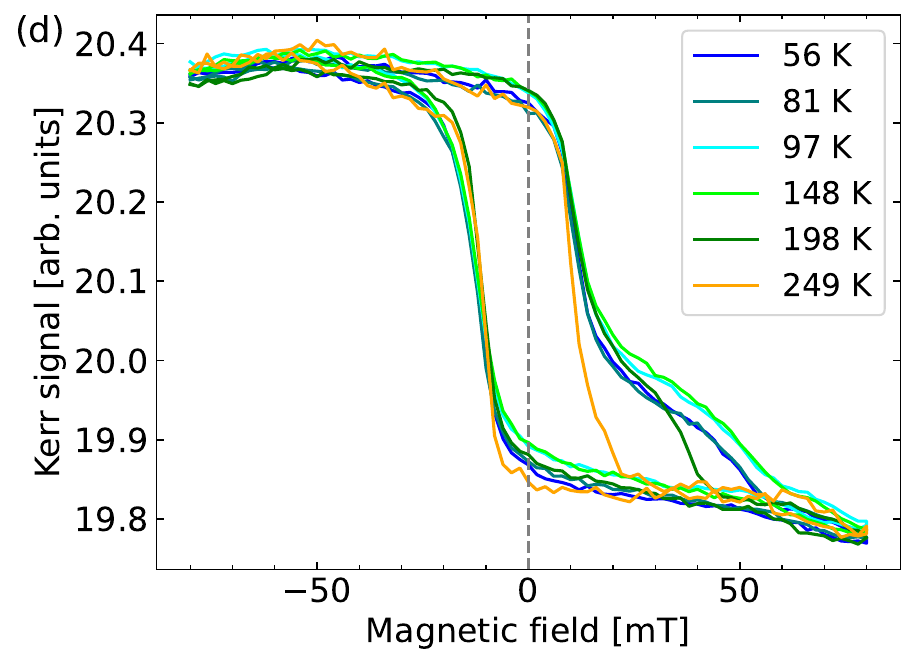}
\includegraphics[width=0.49\columnwidth]{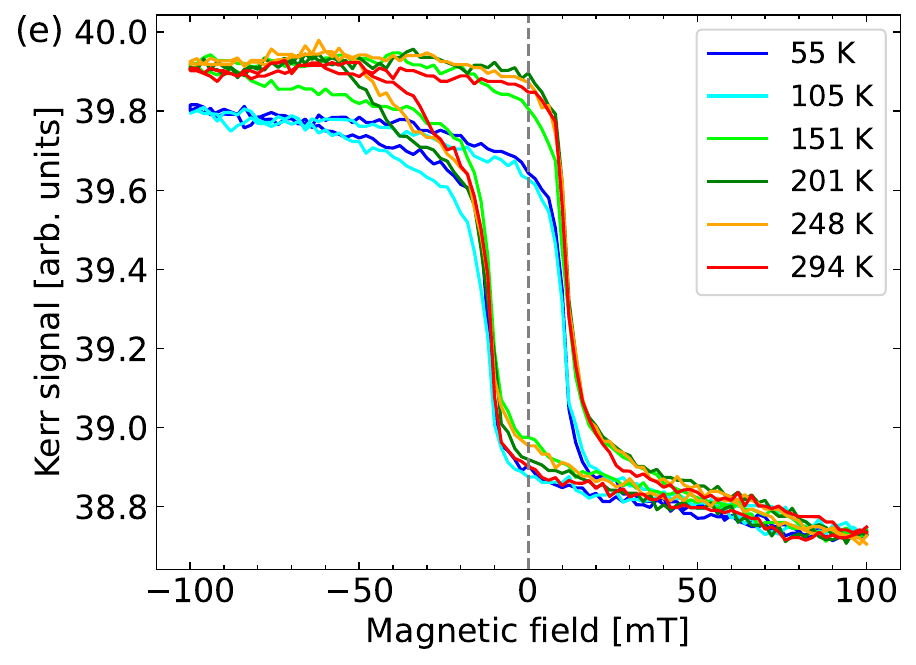}
\includegraphics[width=0.49\columnwidth]{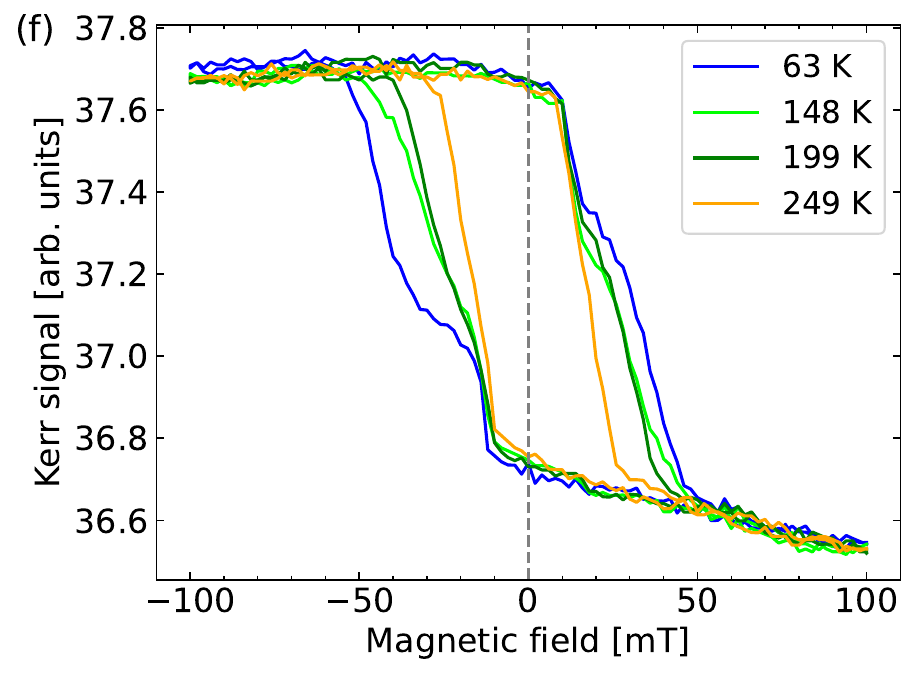}
	\centering
	\caption{(a): Temperature-dependent magnetization loops of a 450\,K postannealed sample after FC. (b): Hysteresis loops after an additional second FC process in reversed field, compared to the initial hysteresis loop at 300\,K after the first FC process. (c-e): Hysteresis loops for different postannealing temperatures (c) 350\,K, (d) 400\,K (FC field applied in negative direction, 17\,ML of $\mathrm{Mn_2Au}$) and (e) 475\,K. (f): Hysteresis loops after FC for a sample with only 12\,ML of $\mathrm{Mn_2Au}$, postannealed at 350\,K.}	
	\label{Fe thickness and FC reversal}
\end{figure}

These areas could depend on the interface roughness between Fe and $\mathrm{Mn_2Au}$, with a flatter interface resulting in a larger coupling area. To investigate this, another sample was heated up in a postannealing step to 450\,K for 30\,min after $\mathrm{Mn_2Au}$ evaporation but before Fe deposition. This should reduce the surface roughness of the $\mathrm{Mn_2Au}$ layer before Fe evaporation and, consequently, reduce the interface roughness. Fig.\ \ref{Fe thickness and FC reversal} (a) shows the temperature-dependent magnetization loops of the 450\,K postannealed sample after FC from 400\,K to 52\,K. Interestingly, the portion of the shifted part reduced from 0.6 without postannealing to about 0.42, compared to Fig.\ \ref{temperature}(b), meaning that less area of $\mathrm{Mn_2Au}$ and Fe is coupled to each other than without the extra postannealing. Additionally, the second sample was field-cooled again, this time from an increased temperature of 450\,K down to 41\,K with an external magnetic field of 100\,mT in the opposite direction to the first FC step. As a result, the shifted part of the hysteresis loops in Fig.\ \ref{Fe thickness and FC reversal} (b) changes its sign of EB from negative to positive. The ratio is again the same between shifted and non-shifted part, so the coupling and non-coupling areas are the same as before. Therefore, increasing the field-cooling temperature seems to have no effect, indicating that the EB can already be set at 400\,K for all coupling areas. Nevertheless, this demonstrates that the shifted part is indeed an EB shift the direction of which can be reset by another FC process.

\begin{figure}[htbp]
\includegraphics[width=0.6\columnwidth]{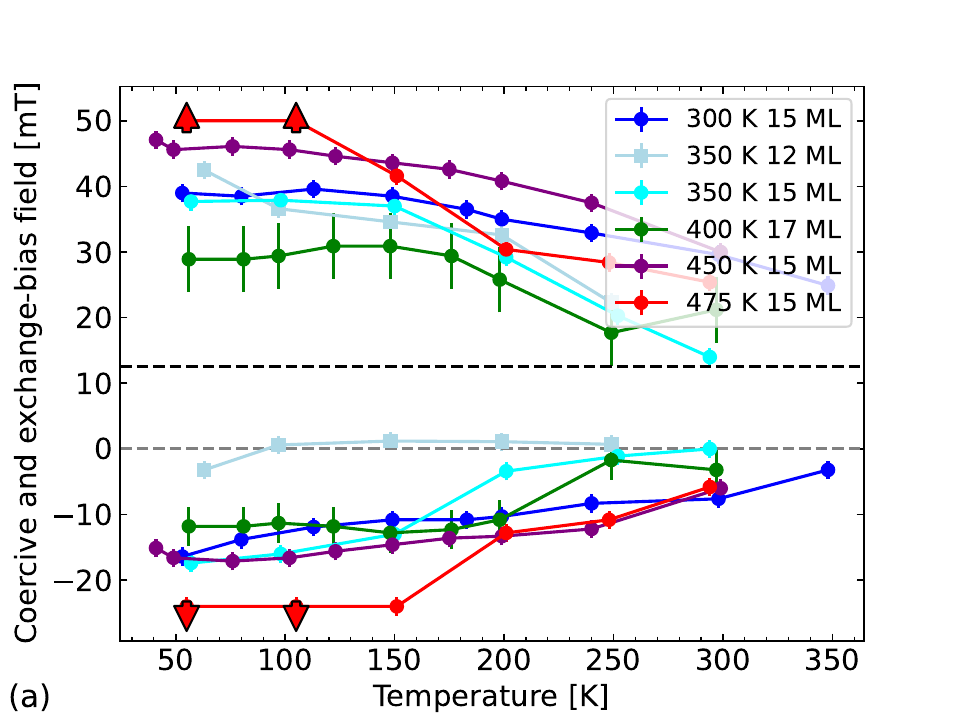}
\includegraphics[width=0.6\columnwidth]{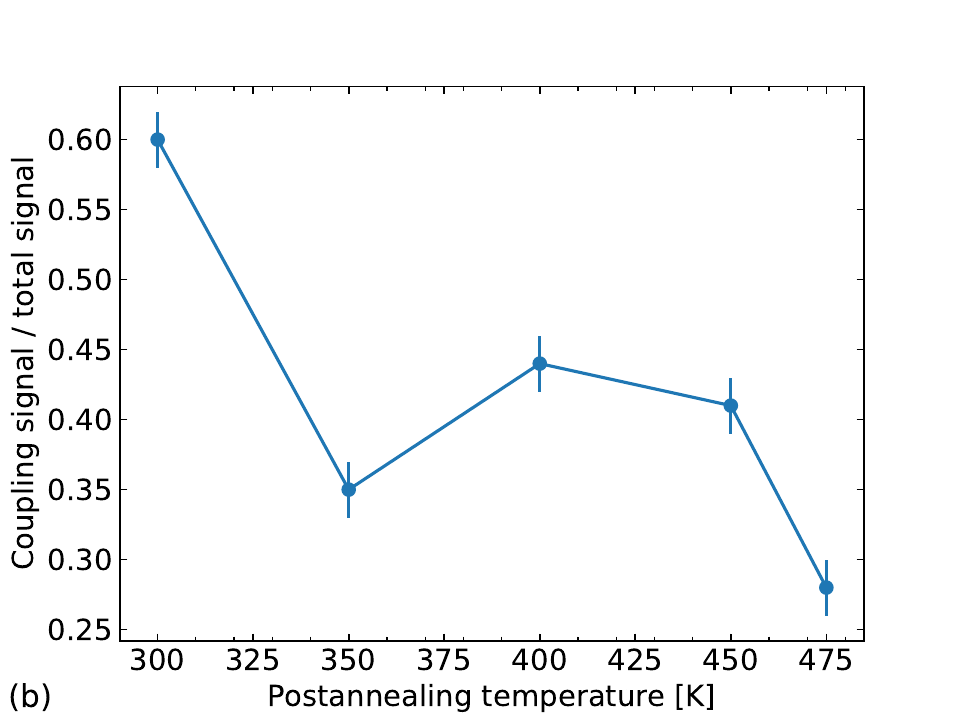}
	\centering
	\caption{(a) Plot of the temperature dependence of the coercive field (top) and EB shift (bottom), taken after a FC process, for different postannealing temperatures ranging from 300\,K to 475\,K. The black dashed horizontal line at 12.5\,mT indicates the coercivity of the non-shifted part. (b) Ratio between the Kerr signal amplitude of the exchange-bias-shifted part of the hysteresis loop and the total amplitude of the saturated hysteresis loops for samples with 15--17\,ML of $\mathrm{Mn_2Au}$.}	
	\label{Postanneal dependence plot}
\end{figure}

To better understand the effect of the postannealing step after $\mathrm{Mn_2Au}$ evaporation, we studied several samples with different postannealing temperatures ranging from room temperature (300\,K) up to 475\,K. Temperature-dependent hysteresis loops of samples with a postannealing temperature of 350\,K, 400\,K and 475\,K are plotted in Fig.\ \ref{Fe thickness and FC reversal} (c-e). Higher postannealing temperatures were not studied, since annealing at or above 500\,K shows a noticeable change in Mn:Au ratio in the AES measurements (with an increased Mn portion due to diffusion), which has also been discussed in literature \cite{Bommanaboyena.2020,Gelen.2024}. Additionally, samples with 12 and 17\,ML (and 20\,ML) of $\mathrm{Mn_2Au}$ were grown to study the influence of $\mathrm{Mn_2Au}$ thickness. Fig. \ref{Postanneal dependence plot} (a) plots the temperature-dependent coercivity and EB shift after FC for different postannealing temperatures, measured while heating up.  All samples show a two-step remagnetization with a temperature-dependent coercivity and EB shift. The 350\,K postannealed sample with only 12\,ML of $\mathrm{Mn_2Au}$ (Fig.\ \ref{Fe thickness and FC reversal} (f)) shows a small EB shift below 100\,K as well as an increased coercivity below 249\,K in both directions, indicating a reset of the EB during the hysteresis loop measurement. The sample with 15\,ML of $\mathrm{Mn_2Au}$ at the same postannealing temperature shows an EB shift also at larger temperatures, up to 201\,K, but also a small increase in coercivity in the positive field direction, which is not present at lower temperatures, indicating a partial reset of the EB. At 252\,K, there is no EB shift but still an increased coercivity while at 300\,K only a single-step hysteresis remains. The other samples with a larger $\mathrm{Mn_2Au}$ thickness still have an EB shift below 300\,K or at even higher temperatures. Therefore, increasing the $\mathrm{Mn_2Au}$ thickness results in an increased blocking temperature. For 20\,ML of $\mathrm{Mn_2Au}$, no EB shift but a temperature dependence of the increased coercivity can be observed (see Supplemental Material \cite{Supplement.2026}). This indicates that field cooling at 400\,K was not sufficient to reach the blocking temperature in that sample. In general, samples with a postannealing temperature of 450\,K and above show a trend towards larger values of EB shift and coercivity. It should be noted that for the sample with a 475\,K postannealing temperature only a lower limit for the coercivity and the absolute EB shift can be determined below 150\,K, indicated by arrows in Fig.\ \ref{Postanneal dependence plot} (a), since an external field of 100\,mT does not saturate the sample. 

The relative coupling area of the sample, i.e., the ratio between the amplitude of the shifted part of the hysteresis loop and its total amplitude (sum of shifted and non-shifted part), is plotted in Fig. \ref{Postanneal dependence plot} (b). Here it seems that the increased EB shift and coercivity comes with a reduction of coupling area, as the amplitude of the second step reduces with larger postannealing temperatures. Sapozhnik \textit{et al}. \cite{Sapozhnik.2019} observed something similar for their sputter-deposited samples, where they explain their findings by different crystallite sizes depending on sample temperature during sputtering. 

\subsection*{Kerr Microscopy}
\begin{figure}[htbp]
\includegraphics[width=0.49\columnwidth]{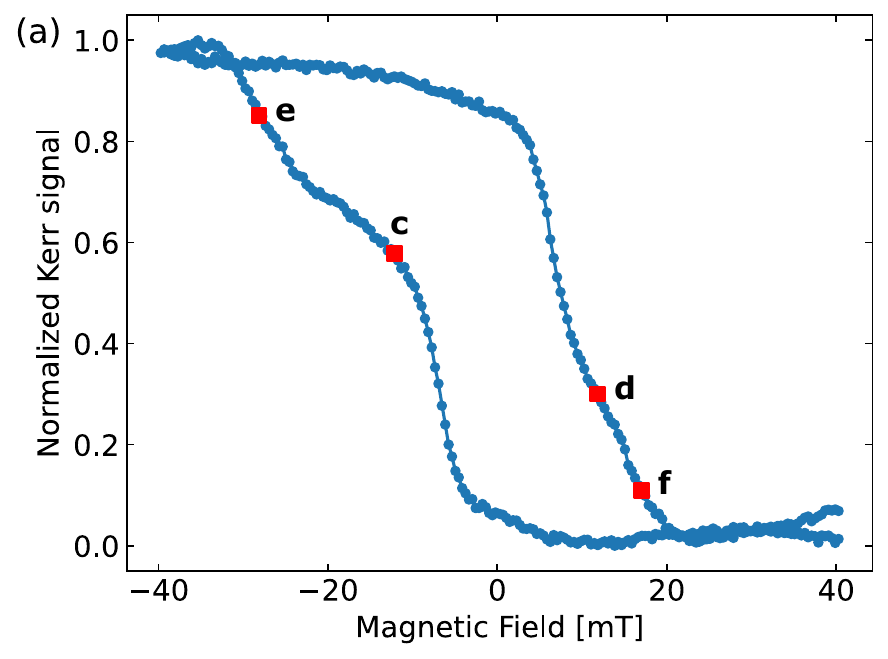}
\includegraphics[width=0.49\columnwidth]{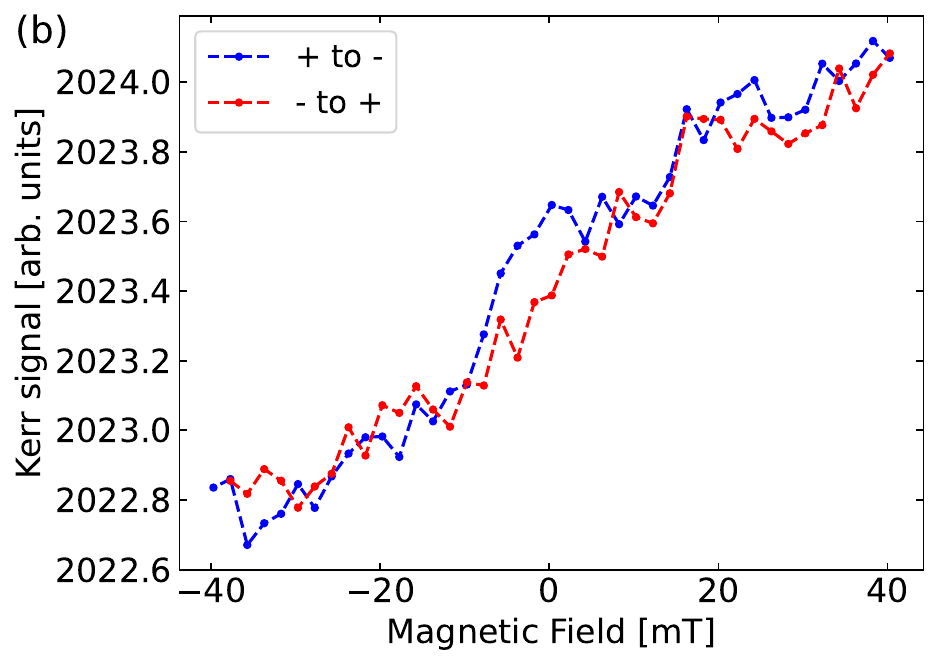}
\includegraphics[width=1.00\columnwidth]{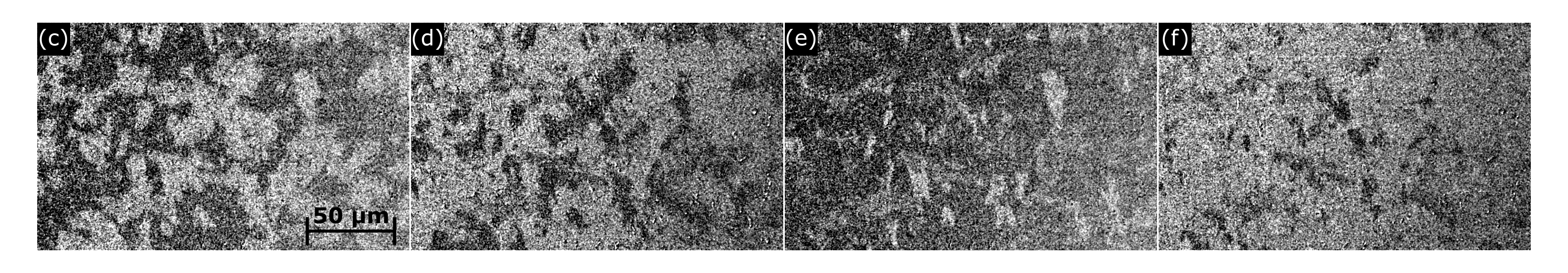}
	\centering
	\caption{Hysteresis loop measured by Kerr microscopy in longitudinal (a) (0.4\,mT steps) and transversal (b) (2\,mT steps) geometry. (c-f): Background-subtracted and drift-corrected longitudinal Kerr microscopy images taken during magnetization reversal for different external magnetic fields: after the first remagnetization step at -12.4\,mT (c) and (d) 11.6\,mT as well as close to saturation at -28.2 mT (e) and 17.2 mT (f), respectively.}	
	\label{Kerr Microscopy}
\end{figure}

To get a closer look at the coupling and non-coupling areas of our 450\,K postannealed sample, it was capped with 15\,ML of Cu to prevent oxidation and imaged by Kerr microscopy in air. Fig.\ \ref{Kerr Microscopy} (a) shows the hysteresis loop of a (220 $\times$ 130)\,µm² sample area, measured in longitudinal geometry with 0.4\,mT field steps \cite{Soldatov.2017,Soldatov.2017b}. It shows a two-step magnetization curve, similar to the one taken with the in-situ L-MOKE in UHV. Fig.\ \ref{Kerr Microscopy} (c-f) shows Kerr images taken at different external magnetic fields to access different states during magnetization reversal. Images are background-subtracted by an image taken at positive saturating field. Dark and bright areas indicate an alignment of Fe domains in positive or negative field direction, respectively. Comparing the images at different fields in the hysteresis loop, it can be easily seen by comparing Kerr image Fig.\ \ref{Kerr Microscopy} (c) and Fig.\ \ref{Kerr Microscopy} (d), taken at -12.4\,mT and 11.6\,mT, respectively, that the same magnetic domain patterns arise, meaning the same regions on the sample that switch at small negative fields also switch at small positive fields. These dark areas in (c) and the bright areas in (d) are the uncoupled areas of the sample while the regions that did not switch are the coupled areas. Kerr images Fig.\ \ref{Kerr Microscopy} (e) and Fig.\ \ref{Kerr Microscopy} (f) were taken at -28.2\,mT and 17.2\,mT, respectively, and show the domain patterns when the coupled areas are partially switched. This means that a larger field of additional 10.5 mT is needed to reach the same state in negative field direction which is caused by the exchange-bias. For the non-coupled areas, about the same field strength is needed in both directions to switch their magnetization direction. The size of the domains is surprisingly large, as they are within the µm range. Additionally, the sample was measured at the same position with polar and transversal geometry to identify the magnetization reversal mechanisms of the Fe layer. 

In polar geometry (see Supplemental Material \cite{Supplement.2026}), the signal did not change during the magnetization reversal, meaning the Fe domains do not rotate out of plane. 
Fig. \ref{Kerr Microscopy} (b) shows the signal taken in transversal geometry. A change in signal, beyond the background slope, can only be observed during the magnetization reversal of the non-coupling areas. This means that these domains switch with a coherent rotation in-plane. On the other hand, the mechanism for the switching of the coupling areas seems to be by domain-wall nucleation and propagation. 

\begin{figure}[htbp]
\includegraphics[width=1.00\columnwidth]{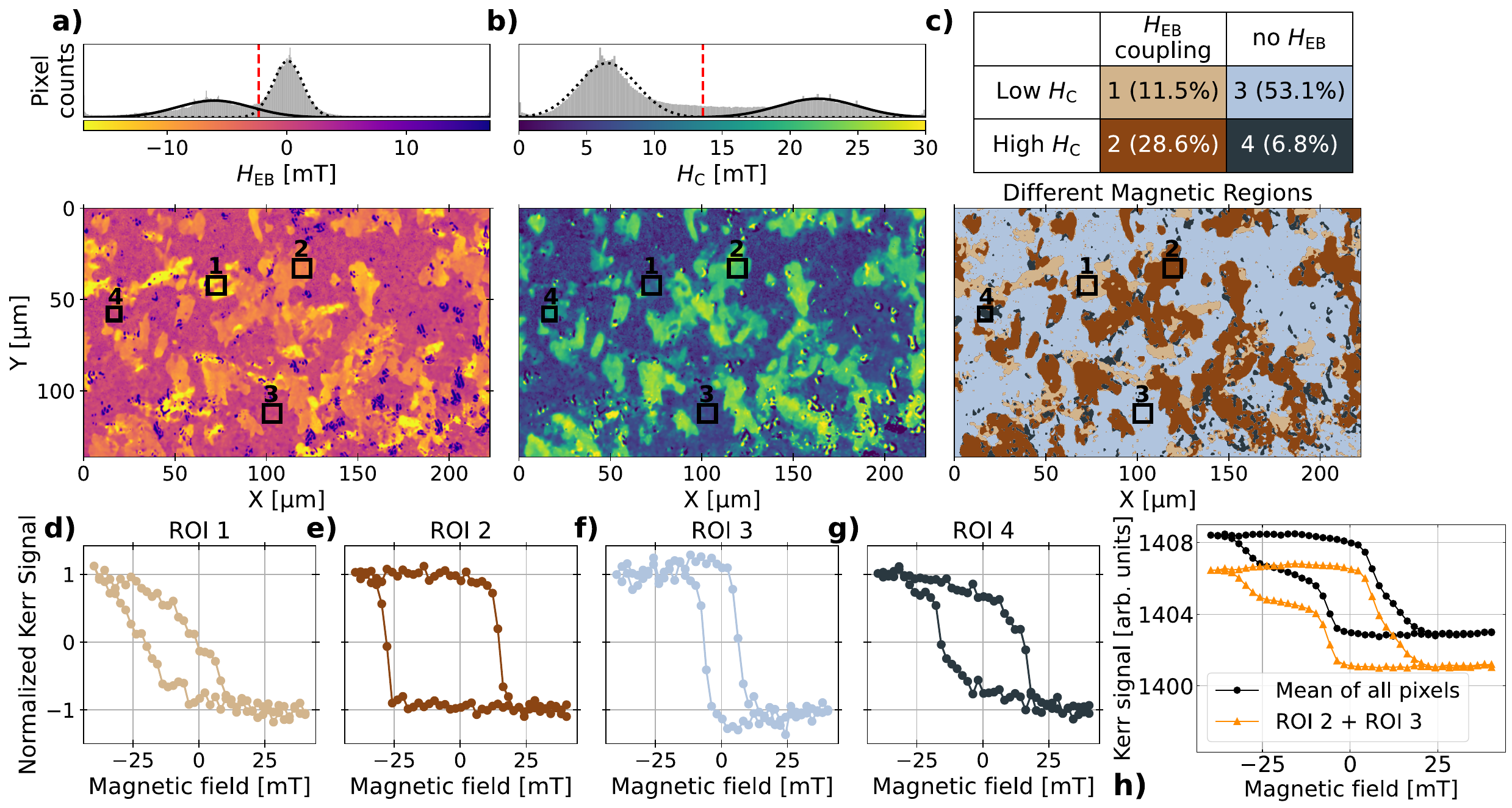}
	\centering
	\caption{Pixel-by-pixel analysis of Kerr microscope images (2\,mT steps). Each pixel's magnetization loop was fitted with an EB-shifted magnetization curve based on arctan functions ($y_\pm = a + (2 / \pi) b \cdot \arctan(c \cdot (x - d \pm e))$). (a) and (b): Color map and corresponding histogram of the local, pixel-wise EB shift ($H\mathrm{_{EB}}$) and coercivity ($H\mathrm{_C}$). The two peaks within both histograms, corresponding to coupled and non-coupled areas of the sample, are fitted by Gaussians (black solid and dotted lines). (c) Depending on the value of $H\mathrm{_C}$ and the absolute value of $H\mathrm{_{EB}}$, four magnetic regions of interest (ROI) can be identified, which are colored in light brown, dark brown, light blue and dark blue. The table shows the relative amount of each ROI within the image. Example areas for each ROI are marked by a black square (1-4) within the images of (a-c). (d-g) Plot of the normalized average Kerr signal as a function of the applied magnetic field for each square. The light brown ROI corresponds to magnetically coupled sample areas that are not saturated by the magnetic field, leading to larger $H\mathrm{_{EB}}$ and smaller $H\mathrm{_C}$ values. Dark brown and light blue ROIs correspond to coupled and non-coupled areas, respectively, while the dark blue ROI represents areas where either the fitting function failed to accurately model the data, returning unreasonably large values, or where the sample exhibits a magnetic coupling without an EB, resulting in a large value of $H\mathrm{_C}$ without an EB shift. (h): Reconstructed loops for the mean Kerr signal over all pixels (black) and from pixels identified as regions of interest (ROI) 2 and 3 (orange).}	
	\label{Pixel analysis}
\end{figure}

In the next step, Kerr images (2\,mT steps) were quantitatively analyzed by fitting the magnetization loop of each pixel with an EB-shifted magnetization curve based on two branches of horizontally shifted arctan functions, one for each step of the magnetization reversal, with the equation $y_\pm = a + (2 / \pi) b \cdot \arctan(c \cdot (x - d \pm e))$. Here, $a$ is a vertical shift in case there are differences in the average Kerr signal between pixels, $b$ is the amplitude, $c$ describes the steepness of the magnetization reversal step, $d$ is the horizontal shift, that means $H\mathrm{_{EB}}$, of the hysteresis loop, while $e$ is a branch-dependent shift with its absolute value being $H\mathrm{_{C}}$. The intensity at each pixel position was averaged by a Gaussian with a FWHM of six pixels to reduce noise. The Gaussian filter represents a compromise between achieving a sufficient signal-to-noise ratio for the fits without losing too much local information. The resulting color map and histogram for the EB shift $H\mathrm{_{EB}}$ and the coercivity $H\mathrm{_C}$ are depicted in Fig.\ \ref{Pixel analysis} (a) and (b). In Fig.\ \ref{Pixel analysis} (c), four regions of interest (ROI) (light brown, dark brown, light blue and dark blue) within the Kerr image can be identified that exhibit a different behavior of the magnetization reversal and are marked by black squares (1-4) for illustration. For each black square the corresponding average Kerr signal is normalized and plotted as a function of the applied magnetic field in Fig.\ \ref{Pixel analysis} (d)-(g). Dark brown (ROI 2) and light blue (ROI 3) regions belong to coupled and non-coupled areas, respectively. The dark brown region exhibits a single-step EB-shifted hysteresis loop with large $H\mathrm{_C}$, while the light blue region shows a non-shifted hysteresis loop with small $H\mathrm{_C}$. From the magnetization loop of the light brown region (ROI 1), it is evident that this area is not saturated by the applied field of 40\,mT. As a result, the fit function returns a rather large $H\mathrm{_{EB}}$ and a small $H\mathrm{_C}$. It can be assumed that if these areas were completely saturated, they would also belong to coupled areas exhibiting both a large $H\mathrm{_{EB}}$ and a large $H\mathrm{_C}$. Finally, dark blue regions (ROI 4) show a large $H\mathrm{_C}$ but no $H\mathrm{_{EB}}$. One reason for this behavior could be a noisy hysteresis loop, for which the fit function partially fails to fit properly and returns unreasonable values, which can also be seen as purple spots in Fig.\ \ref{Pixel analysis} (a). The origin for the large noise may be defects or an inhomogenous illumination of the sample. Another reason could be a coupling area without an EB shift, which is marked by black square 4, with the corresponding hysteresis loop plotted in (g). For this region the direction of the EB could not be set, which could be either due to a very large magnetic anisotropy of the $\mathrm{Mn_2Au}$ domain, preventing the FC process from setting the direction, or the magnetic anisotropy of the $\mathrm{Mn_2Au}$ domain is actually small enough that the whole domain follows the Fe during the magnetization reversal, as has been shown in the literature for polycrystalline systems \cite{Mueglich.2016, Merkel.2020}. Since all dark blue regions are rather small, the latter explanation seems to be most likely the case.

Setting aside dark blue regions, it is evident that sample areas exhibiting a large $H\mathrm{_C}$ also exhibit a significant $H\mathrm{_{EB}}$. Additionally, it can be concluded that the two-step hysteresis loop from the L-MOKE measurements is the result from two single hysteresis loops belonging either to coupled or non-coupled areas. The histogram of the color map exhibits two peaks each for $H\mathrm{_{EB}}$ and $H\mathrm{_C}$, corresponding to coupled and non-coupled areas, with the larger peak belonging to the non-coupled areas. Fitting a Gaussian to each peak, as is shown in Fig.\ \ref{Pixel analysis} (a) and (b), and taking the area ratio between the two Gaussian functions results in a non-coupled:coupled ratio of 1.74:1 for the $H\mathrm{_{EB}}$ and a ratio of 1.47:1 for the $H\mathrm{_C}$ histogram. Here, some of the difference may be explained by the dark blue regions exhibiting no EB shift. Not only are these area ratios quite similar, but they are also very close to the 1.41:1 ratio between the two hysteresis loop steps observed in the L-MOKE measurement. Differences may arise from an imperfect Gaussian fit as well as local variations in the amount of coupling areas, since Kerr images and L-MOKE measurements were taken on different sample regions. It should also be noted that the classification into four distinct magnetic ROIs represents a relatively simplistic approach and may not fully capture the complexity of the sample. However, a hysteresis loop was reconstructed using only the pixels from ROI 2 and 3 (orange curve, Fig.\ \ref{Pixel analysis} (h)). This loop exhibits a shape consistent with the loop generated from the mean intensity of all pixels (black curve), confirming the accurate identification of these ROIs. More importantly, the pixel-wise analysis further confirms the connection between $H\mathrm{_{EB}}$ and an increased $H\mathrm{_{C}}$ for coupled regions as well as the relative amounts of coupled and non-coupled regions.

\section{Discussion}
\label{Discussion}
From the dependence on the postannealing temperature, it is clear that the annealing influences the coupling between Fe and $\mathrm{Mn_2Au}$. However, it does not clarify the physical origin behind coupling and non-coupling areas. In the following, we will discuss different reasons for the occurrence of these areas. 

Samples from Wu \textit{et al}. for an Fe/$\mathrm{Mn_2Au}$ bilayer on MgO(001) show a two-step magnetization reversal as well, but they only observe coupling areas, that means both reversal steps are temperature dependent. Additionally, in contrast to our results, the relative amplitude of both steps depends on the Fe thickness with the non-shifted part decreasing and the shifted part increasing for larger Fe thickness. Here, it should be noted that $\mathrm{Mn_2Au}$ is grown on top of Fe and there is a small Au buffer layer on top of Fe, so the Fe-$\mathrm{Mn_2Au}$ interface is always Au. The asymmetric hysteresis loop is explained to be the result of intergrain exchange coupling and different $\mathrm{Mn_2Au}$ domain configurations. Otherwise, the non-shifted part should not reduce with temperature.

Since for our samples the non-shifted part of the hysteresis loop shows no significant temperature dependence we can discard different domain configurations within the $\mathrm{Mn_2Au}$ layer as an explanation. Oxidation of the $\mathrm{Mn_2Au}$ layer can be excluded as a cause, since AES measurements showed no oxygen peaks after the growth. Surface roughness of the $\mathrm{Mn_2Au}$ layer changes the interface to the Fe layer and could lead to non-coupling areas by reducing the direct exchange interaction. However, MEED oscillations during growth and clearly visible LEED spots indicate a layer-by-layer growth with small surface roughness and long-range order. Also, Kerr microscopy images show domains for the coupling areas with sizes in the tens-of-µm range, larger than what would be expected from roughness. In addition, increasing the post-anneal temperature would then be expected to result in a larger EB-shifted amplitude, but here it shows the opposite effect. On the contrary, our postannealing temperature dependence would in this case indicate that a smaller amount of surface roughness actually results in a larger coupling area.

One other reason could be the growth of the Fe layer on top of $\mathrm{Mn_2Au}$, which shows no MEED oscillations, since the lattice constant is quite different between both layers. Fe may not grow as a complete layer but instead show an island-like growth where only some areas have a good interface contact between $\mathrm{Mn_2Au}$ and Fe. However, this should not be influenced by the postannealing temperature of the $\mathrm{Mn_2Au}$ layer before Fe evaporation. 

A change in the $\mathrm{Mn_2Au}$ crystallite size by the postannealing step could result in an increase or decrease of the Néel temperature $T_N$ so that less or more crystallites can be oriented along the field cooling direction. These crystallites should, however, still couple to Fe, leading to a non-shifted part of the hysteresis loop with increased coercivity and temperature dependence which we do not observe. 

Another reason could be the interface termination of the $\mathrm{Mn_2Au}$ layer to the Fe layer. During the growth process there will either be Au or Mn atoms on the surface and an incomplete $\mathrm{Mn_2Au}$ layer should have both endings. If we assume that only one species couples to Fe while the other does not couple, then this could lead to the two-step hysteresis loop that we observe. Jourdan \textit{et al}. \cite{Bommanaboyena.2021} observed that a $\mathrm{Mn_2Au}$ layer with Au atoms at the interface couples very well to Py. We can assume that this is also the case for our samples. Diffusion of Mn atoms to the surface after growth would then lead to a reduction of the coupling area. Our AES measurements show a slightly Mn-rich $\mathrm{Mn_2Au}$ layer with a ratio close to 2:1, but with a slightly larger Mn portion. This points towards Mn surface diffusion and is in good agreement to our postannealing temperature dependence, where a higher postannealing temperature could mean more Mn surface diffusion and therefore less coupling area. Agglomeration to Mn-rich and Au-rich areas at the interface would also explain the tens-of-µm-sized domains we observe in our Kerr images. In comparison, Bommanaboyena \textit{et al}. and Saposhnik \textit{et al}. observed an average domain size around one magnitude smaller with a size of 1 to 3 µm for $\mathrm{Mn_2Au}$/Py and pure $\mathrm{Mn_2Au}$ layers grown on a Ta buffer layer, respectively. \cite{Bommanaboyena.2021, Sapozhnik.2018}. 
 
Therefore, we conclude that coupling and non-coupling areas most likely result from a different interface termination of the $\mathrm{Mn_2Au}$ layer, whereby a Au termination results in a magnetic coupling to Fe. Cross-sectional TEM could give a better insight into our $\mathrm{Mn_2Au}$/Fe interface, but would go beyond the scope of this research. Additionally, changing the interface after Fe evaporation, without causing intermixing, would be interesting to carry out in future studies. 

\section{Conclusion}
\label{Conclusion}
In summary, we achieved layer-by-layer growth of a room-temperature AFM $\mathrm{Mn_2Au}$ layer on a metallic Nb(001)-like substrate, extending the range of available substrates and giving more insight into the growth mode of $\mathrm{Mn_2Au}$. Evaporation of Fe on top of $\mathrm{Mn_2Au}$ and subsequent FC results in a partial EB shift, which can be reset by another FC process, as well as a temperature-dependent coercivity for the shifted part. Both effects originate from the coexistence of coupling and non-coupling areas between both layers. Kerr microscopy images reveal the size of the coupling areas to be in the tens-of-µm range. We observed a decreased amount of coupling area for larger postannealing temperatures, which means that interface roughness is not the reason for the non-coupling areas. Instead, we conclude that the interface termination of the $\mathrm{Mn_2Au}$ layer is the cause for the existence of coupling and non-coupling areas. 

Our findings highlight the importance of high-quality $\mathrm{Mn_2Au}$ layer growth but also gives rise to many more questions regarding the interface and the coupling between $\mathrm{Mn_2Au}$ and a ferromagnetic layer. Understanding the coupling behavior is crucial for the application in AFM spintronic devices.

%

\end{document}


\title{Growth and Kerr magnetometry of Mn$_2$Au on a gold-capped Nb(001) substrate 
\\--- Supplemental Material ---}

\author{Jendrik G\"ordes\orcidlink{0000-0003-4321-8133}}
\affiliation{Institut f\"ur Experimentalphysik, Freie Universit\"at Berlin, Arnimallee 14, 14195 Berlin, Germany}
\author{Christian Janzen\orcidlink{0000-0003-0232-2671}}
\affiliation{Institute of Physics and Center for Interdisciplinary Nanostructure Science and Technology (CINSaT), University of Kassel, Heinrich-Plett-Stra{\ss}e 40, 34132 Kassel, Germany}
\author{Arne J. Vereijken\orcidlink{0009-0001-5009-5859}}
\affiliation{Institute of Physics and Center for Interdisciplinary Nanostructure Science and Technology (CINSaT), University of Kassel, Heinrich-Plett-Stra{\ss}e 40, 34132 Kassel, Germany}
\author{Tingwei Li}
\affiliation{Institut f\"ur Experimentalphysik, Freie Universit\"at Berlin, Arnimallee 14, 14195 Berlin, Germany}
\author{Tauqir Shinwari\orcidlink{0000-0002-3876-3201}}
\affiliation{Institut f\"ur Experimentalphysik, Freie Universit\"at Berlin, Arnimallee 14, 14195 Berlin, Germany}
\author{Rico Huhnstock\orcidlink{0000-0002-3326-8084}}
\affiliation{Institute of Physics and Center for Interdisciplinary Nanostructure Science and Technology (CINSaT), University of Kassel, Heinrich-Plett-Stra{\ss}e 40, 34132 Kassel, Germany}
\author{Arno Ehresmann\orcidlink{0000-0002-0981-2289}}
\affiliation{Institute of Physics and Center for Interdisciplinary Nanostructure Science and Technology (CINSaT), University of Kassel, Heinrich-Plett-Stra{\ss}e 40, 34132 Kassel, Germany}
\author{Wolfgang Kuch\orcidlink{0000-0002-5764-4574}}
\email{Correspondence and requests for materials should be addressed to W.K. (email: kuch@physik.fu-berlin.de)}
\affiliation{Institut f\"ur Experimentalphysik, Freie Universit\"at Berlin, Arnimallee 14, 14195 Berlin, Germany}

\date{\today}

\renewcommand{\thefigure}{S\arabic{figure}}
\renewcommand{\theequation}{S\arabic{equation}}
\renewcommand{\thetable}{S \Roman{table}}

\maketitle

\subsection*{Auger electron spectroscopy}
After co-evaporation of Mn and Au, Auger electron spectroscopy was employed to determine the composition of the alloy on the sample surface. For our Auger system, parameters used for calculating the stoichiometric composition of $\mathrm{Mn_{1-x}Au_{x}}$ are shown in Tab.\ \ref{tab:S1}. The mean free path $\lambda$ of Au at 69 eV was approximated by an empirical relation \cite{Memeo.1983}. The sensitivity $S$ of Au was experimentally determined from a $\mathrm{Cu_3Au}$ single crystal, while for Mn, $S$ and $\lambda$ were determined by a series of grown thin-film layers with known thickness. Both $S$ values were calibrated in relation to the 920\,eV peak of a pure Cu substrate with $S_\mathrm{Cu}$ set to 1. With the Auger peak-to-peak intensities of Au $I_{\mathrm{Au}}$ and Mn $I_{\mathrm{Mn}}$ the stoichiometry is calculated by
\begin{equation}
\frac{I_{\mathrm{Au}}}{I_{\mathrm{Mn}}} = \frac{S_{\mathrm{Au}}}{S_{\mathrm{Mn}}}~\frac{1-e^{-d/\lambda_{\mathrm{Au}}}}{1-e^{-d/\lambda_{\mathrm{Mn}}}}~\frac{x}{1-x}
\end{equation}
with $d$ being the layer thickness of $\mathrm{Mn_{1-x}Au_{x}}$.

\begin{table}[htbp]
  \centering
  \caption{Auger constants used for the calculation of the Au:Mn ratio.}
    \begin{tabular}{|c|c|c|c|}
    \hline
    Element & Energy [eV] & Sensitivity $S$ & Mean free path $\lambda$ [ML] \\
    \hline
    Au & 69 & 2.13 ($\pm$ 0.09) & 1.88 ($\pm$ 0.09) \\
    Mn & 589 & 0.88 ($\pm$ 0.06) & 4.0 ($\pm$ 0.2) \\
    \hline
    \end{tabular}
  \label{tab:S1}
\end{table}

\subsection*{Angular polarization dependence of MOKE measurements}
\begin{figure}[htbp]
\includegraphics[width=0.49\columnwidth]{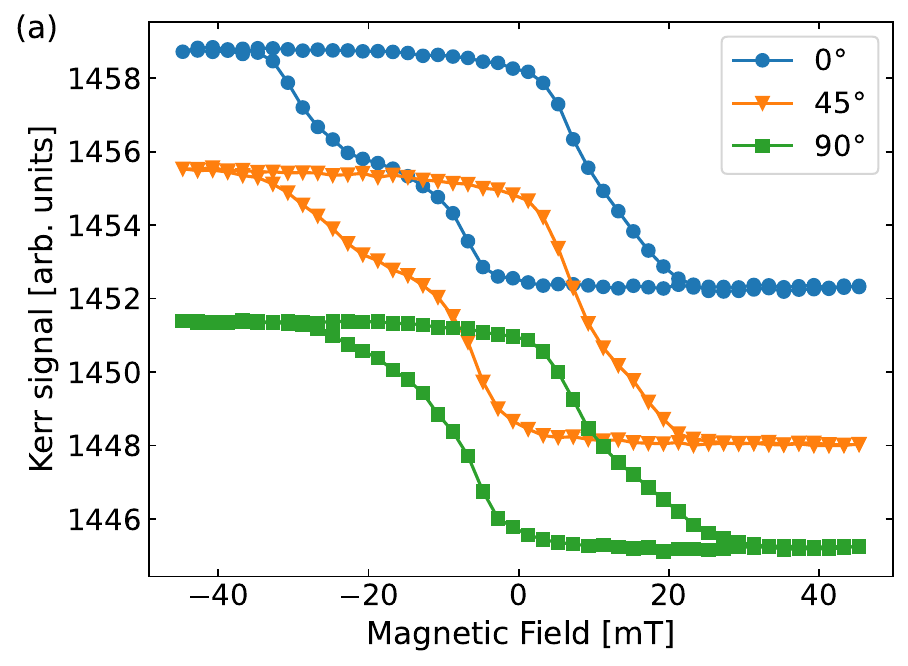}
\includegraphics[width=0.49\columnwidth]{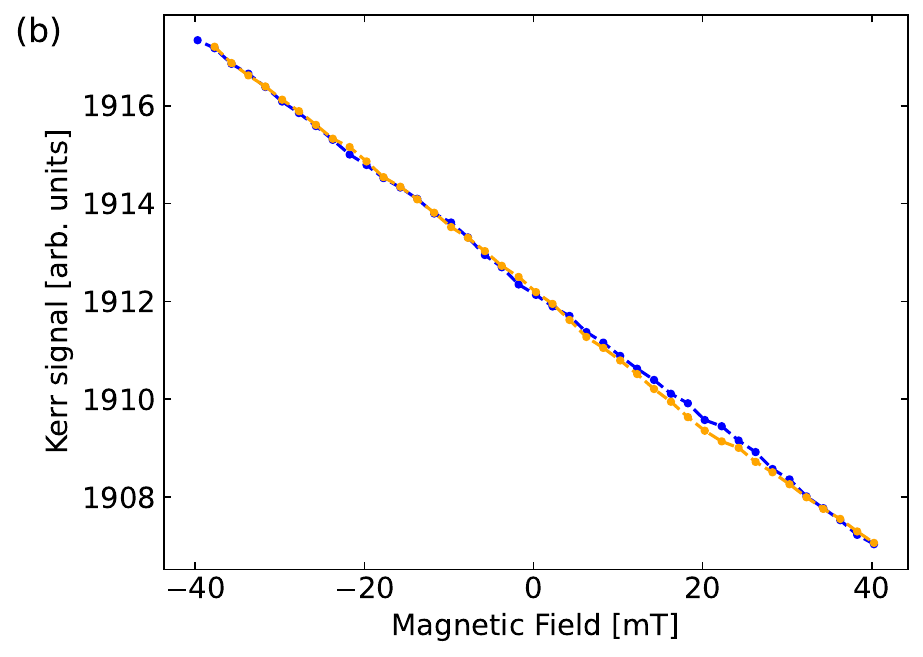}
	\centering
	\caption{Kerr microscopy hysteresis loops measured in longitudinal geometry along (0°), perpendicular (90°) and 45° to the FC direction (a) as well as in polar geometry (b). In (b), the external magnetic field was applied in-plane along the 0°-axis.}	
	\label{Angular dependency}
\end{figure}

A non-postannealed Nb(001)/NbAu/15\,ML $\mathrm{Mn_2Au}$/20\,ML Fe/15\,ML Cu sample was measured by Kerr microscopy in the longitudinal geometry \cite{Soldatov.2017,Soldatov.2017b} and rotated in-plane by 0°, 45° and 90° with respect to the [110]-direction. The hysteresis loops in Fig. \ref{Angular dependency} (a) show two steps for all three angles, however, there is no more EB shift at an azimuthal angle of 90°. The hysteresis loop at an angle of 45° seems to be a mixture of the ones at 0° and 90°. This shows that the magnetization of the Fe layer is indeed aligned along the [110]-direction of $\mathrm{Mn_2Au}$ and that the EB field was set in the direction of the external field from the FC process. There is no EB shift in 90° (and 270°) direction because it is at an angle of 90° to the set FC direction. Changing the Kerr microscope setup to be sensitive for the polar MOKE effect, Fig. \ref{Angular dependency}(b) shows no significant change in Kerr signal during magnetization reversal with the external magnetic field applied in-plane. An out-of-plane rotation during magnetization reversal can therefore be excluded. 

\subsection*{MOKE measurement of a 20 ML Mn$_2$Au sample}
\begin{figure}[htbp]
\includegraphics[width=0.55\columnwidth]{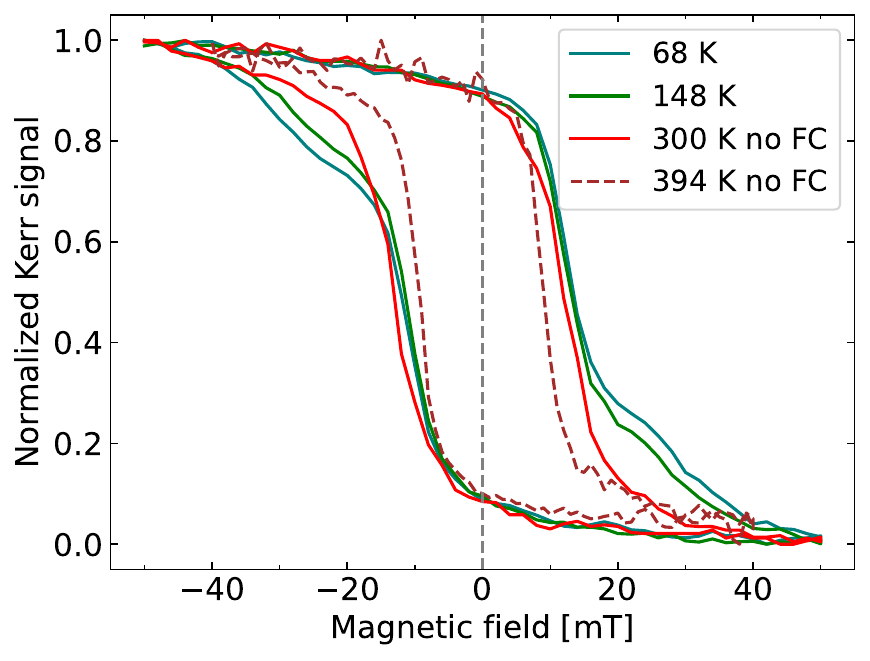}
	\centering
	\caption{Hysteresis loops of a non-postannealed Nb(001)/NbAu/20\,ML $\mathrm{Mn_2Au}$/15\,ML Fe sample before and after a FC process from 400\,K.}	
	\label{20ML}
\end{figure}

To investigate the effect of a thicker $\mathrm{Mn_2Au}$ layer, a not postannealed Nb(001)/NbAu/20 ML $\mathrm{Mn_2Au}$/15 ML Fe sample was measured by L-MOKE. Fig. \ref{20ML} shows the temperature-dependent hysteresis loops before and after FC from 400\,K. A two-step hysteresis loop is observed with no EB shift but an increased coercivity. Since the coercivity is increased in both directions, we conclude that for a 20\,ML thick $\mathrm{Mn_2Au}$ layer a FC temperature of 400\,K is not sufficient to set the direction of the EB. Thus, the Néel-vector orientation of the Mn spins is randomly distributed, contributing to the increased coercivity in both directions. 

%